\documentclass[12pt]{article}

\usepackage{graphicx}
\begin{document}
\begin{center}
{\bf Modified nonlinear model of arcsin-electrodynamics }\\
\vspace{5mm}
 S. I. Kruglov
\footnote{E-mail: serguei.krouglov@utoronto.ca}
 \\

\vspace{5mm}
\textit{Department of Chemical and Physical Sciences, University of Toronto,\\
3359 Mississauga Road North, Mississauga, Ontario, Canada L5L 1C6}
\end{center}

\begin{abstract}
A new modified model of nonlinear arcsin-electrodynamics with two parameters is proposed and
analyzed. We obtain the corrections to the Coulomb law. The effect of vacuum birefringence takes place when the external constant magnetic field is present. We calculate indices of refraction for two perpendicular polarizations of electromagnetic waves and estimate bounds on the parameter $\gamma$ from the BMV and PVLAS experiments. It is shown that the electric field of a point-like charge is finite at the origin. We calculate the finite static electric energy of point-like particles and demonstrate that the electron mass can have the pure electromagnetic nature. The symmetrical Belinfante energy-momentum tensor and dilatation current are found. We show that the dilatation symmetry and dual symmetry are broken in the model suggested. We have investigated the gauge covariant quantization of the nonlinear electrodynamics fields as well as the gauge fixing approach based on Dirac's brackets.
\end{abstract}

\vspace{5mm}



\vspace{5mm}

\section{Introduction}

In Maxwell's electrodynamics a point-like charge possesses an infinite electromagnetic energy but in Born-Infeld (BI) electrodynamics \cite{Born}, \cite{Infeld}, \cite{Plebanski}, where there is a new parameter with the dimension of the length, that problem of singularity is solved. In BI electrodynamics the dimensional parameter gives the maximum of the  electric fields. In addition, nonlinear electrodynamics may give a finite electromagnetic energy of a charged point-like particle. As a result, in such models the electron mass can have pure electromagnetic nature. It is known that in QED one-loop quantum corrections contribute to classical electrodynamics and give non-linear terms in the Lagrangian \cite{Heisenberg}, \cite{Schwinger}, \cite{Adler}. Different models of non-linear electrodynamics were investigated in \cite{Kruglov}, \cite{Kruglov3}, \cite{Hendi}, \cite{Kruglov2} and  \cite{Kruglov1}.
In this paper we analyze the new model with two parameters which is the modification of the model of nonlinear arcsin-electrodynamics considered in \cite{Kruglov5}.
The nonlinear effects should be taken into account for strong electromagnetic fields.

The structure of the paper is as follows. In section 2, we propose the Lagrangian of new model of nonlinear electrodynamics. The field equations are written in the form of Maxwell's equations where
the electric permittivity $\varepsilon_{ij}$ and  magnetic permeability $\mu_{ij}$ tensors depend on electromagnetic fields. In section 3 we show that the electric field of a point-like charge is not singular at the origin and have the finite value. We find the correction to the Coulomb law. The phenomenon of vacuum birefringence is investigated in section 4 and we estimate the bounds on the parameter $\gamma$ from BMV and PVLAS experiments. In  section 5 we obtain the symmetrical Belinfante energy-momentum tensor, the dilatation current, and its non-zero divergence. The finite static electric energy of point-like particles is calculated and we demonstrate that the electron mass can be treated as the pure electromagnetic energy. The gauge covariant quantization of the nonlinear electrodynamics proposed was studied in Sec. 6. The gauge fixing approach based on Dirac's brackets is investigated in subsection 6.1.
We discuss the result obtained in section 7.

The Lorentz-Heaviside units with $\hbar =c=\varepsilon_0=\mu_0=1$ and the Euclidian metrics are used. Greek
letters run from $1$ to $4$ and Latin letters run from $1$ to $3$.

\section{Field equations of the model}

We propose nonlinear electrodynamics with the Lagrangian density
\begin{equation}
{\cal L} = -\frac{1}{\beta}\arcsin\left(\beta{\cal F}-\frac{\beta\gamma}{2}{\cal G}^2\right),
 \label{1}
\end{equation}
where $\beta$, $\gamma$ are dimensional parameters ($\beta{\cal F}$ and $\beta\gamma{\cal G}^2$ are dimensionless). The Lorentz-invariants are defined by ${\cal F}=(1/4)F_{\mu\nu}^2=(\textbf{B}^2-\textbf{E}^2)/2$, ${\cal G}=(1/4)F_{\mu\nu}\tilde{F}_{\mu\nu}=\textbf{E}\cdot \textbf{B}$, and $F_{\mu\nu}=\partial_\mu A_\nu-\partial_\nu A_\mu$ is the field strength,  $\tilde{F}_{\mu\nu}=(1/2)\varepsilon_{\mu\nu\alpha\beta}F_{\alpha\beta}$ is a dual tensor ($\varepsilon_{1234}=-i$) and $A_\mu$ is the 4-vector-potential.
The nonlinear model of electrodynamics introduced can be considered as an effective model. At weak electromagnetic fields the model based on the Lagrangian density (1) approaches to classical electrodynamics, ${\cal L}\simeq -{\cal F}$, and the correspondence principle takes place \cite{Shabad}. The variables $\beta^{1/4}$, $\gamma^{1/4}$ possess the dimension of the length and can be considered as fundamental constants in the model.

Euler-Lagrange equations lead to the equations of motion
\begin{equation}
\partial_\mu\left(\frac{F_{\mu\nu}-\gamma {\cal G}\tilde{F}_{\mu\nu}}{\sqrt{1-\left(\beta{\cal F}-\frac{\beta\gamma}{2}{\cal G}^2\right)^2}}\right)=0.
\label{2}
\end{equation}
The electric displacement field,
$\textbf{D}=\partial{\cal L}/\partial \textbf{E}$ ($E_j=iF_{j4}$), is given by
\begin{equation}
\textbf{D}=\frac{1}{\Pi}\left(\textbf{E}+\gamma{\cal G}\textbf{B}\right),~~~~\Pi=\sqrt{1-\left(\beta{\cal F}-\frac{\beta\gamma}{2}{\cal G}^2\right)^2}.
\label{3}
\end{equation}
The magnetic field can be found from the relation $\textbf{H}=-\partial{\cal L}/\partial \textbf{B}$ ($B_j=(1/2)\varepsilon_{jik}F_{ik}$, $\varepsilon_{123}=1$),
\begin{equation}
\textbf{H}= \frac{1}{\Pi}\left(\textbf{B}-\gamma{\cal G}\textbf{E}\right).
\label{4}
\end{equation}
One can decompose Eqs. (3), (4) as follows \cite{Hehl}:
\begin{equation}
D_i=\hat{\varepsilon}_{ij}E_j+\nu_{ij}B_j,~~~~H_i=(\hat{\mu}^{-1})_{ij}B_j-\nu_{ji}E_j.
\label{5}
\end{equation}
From Eqs. (3), (4), ( 5) we obtain
\begin{equation}
\hat{\varepsilon}_{ij}=\delta_{ij}\frac{1}{\Pi},~~~~(\hat{\mu}^{-1})_{ij}=\delta_{ij}\frac{1}{\Pi},~~~~\nu_{ji}=\delta_{ij}\frac{\gamma {\cal G}}{\Pi}.
\label{6}
\end{equation}
It is more convenient for our model to define the displacement field and the magnetic induction field as \cite{Dittrich}  $D_i=\varepsilon_{ij}E_j$, $B_i=\mu_{ij}H_j$. Then we find the electric permittivity tensor $\varepsilon_{ij}$ and the inverse magnetic permeability tensor $(\mu^{-1})_{ij}$ from Eqs. (3), (4),
\begin{equation}
\varepsilon_{ij}=\frac{1}{\Pi}\left(\delta_{ij}+\gamma B_iB_j\right).
\label{7}
\end{equation}
\begin{equation}
(\mu^{-1})_{ij}=\frac{1}{\Pi}\left(\delta_{ij}-\gamma E_iE_j\right).
\label{8}
\end{equation}
Thus, instead of three tensors (6) we introduce only two symmetrical tensors (7), (8). It should be noted that both decompositions lead to the same field equations (2), and therefore they are equivalent.
Equations of motion (2) may be written, with the help of Eqs. (3),(4), in the form of the first pair of the Maxwell equations
\begin{equation}
\nabla\cdot \textbf{D}= 0,~~~~ \frac{\partial\textbf{D}}{\partial
t}-\nabla\times\textbf{H}=0.
\label{9}
\end{equation}
The second pair of Maxwell's equations follows from the Bianchi identity
and is given by
\begin{equation}
\nabla\cdot \textbf{B}= 0,~~~~ \frac{\partial\textbf{B}}{\partial
t}+\nabla\times\textbf{E}=0.
\label{10}
\end{equation}
Eqs. (9), (10) are nonlinear Maxwell equations because $\varepsilon_{ij}$ and $\mu_{ij}$ depend on the electromagnetic fields $\textbf{E}$ and $\textbf{B}$.
From Eqs. (3),(4) we obtain the relation
\begin{equation}
\textbf{D}\cdot\textbf{H}=\frac{\textbf{E}\cdot\textbf{B}}{\Pi^2}\left(1+2\gamma{\cal F}-\gamma^2{\cal G}^2\right).
\label{11}
\end{equation}
The dual symmetry is broken in this model because $\textbf{D}\cdot\textbf{H}\neq\textbf{E}\cdot\textbf{B}$ \cite{Gibbons}. It should be noted that BI electrodynamics is dual symmetrical but in generalized BI electrodynamics \cite{Kruglov4} the dual symmetry is broken as well as in QED due to one loop quantum corrections.

\section{The field of the point-like charged particles}

From Maxwell's equation (9) in the presence of the point-like charge source, we obtain the equation
\begin{equation}
\nabla\cdot \textbf{D}=e\delta(\textbf{r})
\label{12}
\end{equation}
having the solution
\begin{equation}
\textbf{D}=\frac{e}{4\pi r^3}\textbf{r}.
\label{13}
\end{equation}
Equation (13) with the help of Eq. (7), and at $\textbf{B}=0$, becomes
\begin{equation}
E\left(\frac{1}{\sqrt{1-\beta^2 E^4/4}}\right)=\frac{e}{4\pi r^2}.
\label{14}
\end{equation}
When $r\rightarrow 0$ the solution to Eq. (14) is
\begin{equation}
E(0)=\sqrt{\frac{2}{\beta}}.
\label{15}
\end{equation}
Equation (15) represents the maximum electric field at the origin of the charged point-like particle. This attribute  is similar to BI electrodynamics, and is in contrast to linear electrodynamics where the electric field strength possesses the singularity.
Let us to introduce unitless variables
\begin{equation}
u=\frac{4\pi r^2}{e\sqrt{\beta}},~~~~v=\sqrt{\frac{\beta}{2}}E.
\label{16}
\end{equation}
Equation (14) using (16) becomes $v^4+2u^2v^2-1=0$ with the real solution
\begin{equation}
v=\sqrt{\sqrt{u^4+1}-u^2}.
\label{17}
\end{equation}
The plot of the function $v(u)$ is given in Fig. 1.
\begin{figure}[h]
\includegraphics[height=3.0in,width=3.0in]{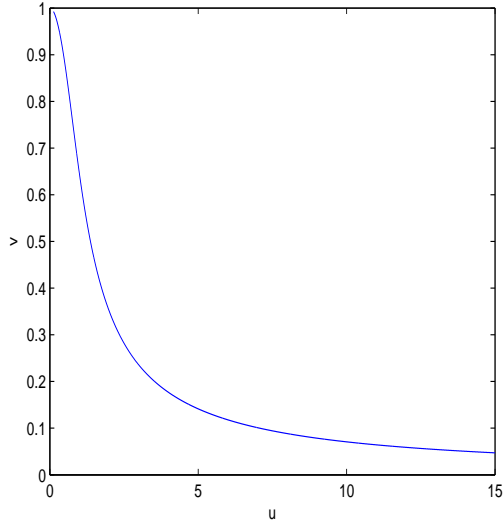}
\caption{\label{fig.1}The function  $v$ vs. $u$.}
\end{figure}
From Eqs. (16), (17) we find the function $E(r)$
\begin{equation}
E(r)=\frac{4\sqrt{2}\pi}{e\beta}\sqrt{\sqrt{r^8+\frac{e^4\beta^2}{(4\pi)^4}}-r^4}.
\label{18}
\end{equation}
One obtains from Eq. (18) the finite value $E(0)=\sqrt{2/\beta}$ at the origin, Eq. (15). Thus, there is no singularity of the electric field at the origin of the point-like charged particles. The Taylor series of the function (18) at $r\rightarrow 0$ is
\begin{equation}
E(r)=\sqrt{\frac{2}{\beta}}\left(1-\frac{8\pi^2r^4}{e^2\beta}+{\cal O}(r^8)\right).
\label{19}
\end{equation}
The expansion of the function (17), $v(u)$, at $u\rightarrow\infty$ is given by
\begin{equation}
v=\frac{1}{\sqrt{2}u}-\frac{1}{8\sqrt{2}u^{5}}+\frac{7}{128\sqrt{2}u^9}+{\cal O}(u^{-13}).
\label{20}
\end{equation}
From Eqs (16), (20) we obtain the asymptotic value of the electric field at $r\rightarrow \infty$
\begin{equation}
E(r)=\frac{e}{4\pi r^2}-\frac{e^5\beta^2}{8(4\pi)^5r^{10}}+{\cal O}(r^{-18}).
\label{21}
\end{equation}
The first term in Eq. (21) gives Coulomb's law and the second term leads to the correction of the Coulomb law at $r\rightarrow\infty$. At $\beta=0$ one comes to Maxwell's electrodynamics and the Coulomb law $E=e/(4\pi r^2)$ is recovered. It follows from Eq. (21) that corrections to the Coulomb law at $r\rightarrow\infty$ are very small. After integration of Eq. (21) we obtain the asymptotic value of the electric potential at $r\rightarrow\infty$
\begin{equation}
A_0(r)=-\frac{e}{4\pi r}+\frac{e^5\beta^2}{72(4\pi)^5r^9}+{\cal O}(r^{-17}).
\label{22}
\end{equation}
As a result, the electric field is finite at the origin of the charged object and there are not singularities of the electric field at the origin of point-like charged objects. It should be noted that the similar situation takes place in BI electrodynamics.

\section{Vacuum birefringence}

It is known that vacuum birefringence takes place in QED due to one-loop quantum corrections \cite{Adler1}, \cite{Biswas}. The effect of birefringence is absent in BI electrodynamics.
We note that in generalized BI electrodynamics \cite{Kruglov4} there is the effect of birefringence. Here we consider the superposition of the external constant and uniform magnetic induction field $\textbf{B}_0=B_0(1,0,0)$ and the plane electromagnetic wave $(\textbf{e}, \textbf{b})$,
\begin{equation}
\textbf{e}=\textbf{e}_0\exp\left[-i\left(\omega t-kz\right)\right],~~~\textbf{b}=\textbf{b}_0\exp\left[-i\left(\omega t-kz\right)\right]
\label{23}
\end{equation}
propagating in $z$-direction, where $k=2\pi/\lambda$ ($\lambda$ is the wave length).
Let us consider the strong magnetic induction field $\textbf{B}$ so that the resultant electromagnetic fields are given by $\textbf{E}=\textbf{e}$, $\textbf{B}=\textbf{b}+\textbf{B}_0$. We assume that amplitudes of the electromagnetic wave, $e_0, b_0$, are much less than the external magnetic induction field, $e_0,b_0\ll B_0$. The approximate Lagrangian density (1) up to ${\cal O}(e_0^2)$, ${\cal O}(b_0^2)$ becomes
\begin{equation}
{\cal L} \approx -\frac{1}{\beta}\arcsin\left\{\frac{\beta}{2}\left[\left(\textbf{B}_0+\textbf{b}\right)^2-\textbf{e}^2\right]- \frac{\beta\gamma}{2}(\textbf{e}\cdot\textbf{B}_0)^2\right\}.
 \label{24}
\end{equation}
We define the electric displacement field \cite{Dittrich} $d_i=\partial{\cal L}/\partial e_i$ and the magnetic field $h_i=-\partial{\cal L}/\partial b_i$, and
linearize equations with respect to the wave fields $\textbf{e}$ and $\textbf{b}$. Implying that $\beta^2 B_0^4\ll 1$ the electric permittivity and magnetic permeability tensor elements are given by
\[
\varepsilon_{ij}=\varepsilon\left(\delta_{ij}+\gamma B_{0i}B_{0j}\right),~~~~
(\mu^{-1})_{ij}=\varepsilon\left(\delta_{ij}+\frac{\varepsilon^2\beta^2B_0^2B_{0i}B_{0j}}{2}\right),
\]
\begin{equation}
\varepsilon=\frac{1}{\sqrt{1-\beta^2B_0^4/4}},
\label{25}
\end{equation}
so that $d_i=\varepsilon_{ij}e_i$, $b_i=\mu_{ij}h_i$. The terms  containing $e_0^2$, $b_0^2$, $e_0b_0$ are small compared with $B_0^2$ and were neglected in Eqs.(25).
We leave only linear terms in $e$, $b$ in expressions for the electric $d_i$ and the magnetic $b_i$ fields.

The diagonal electric permittivity and magnetic permeability tensor elements are
\[
\varepsilon_{11}=\varepsilon\left(1+\gamma B_0^2\right),~~~~\varepsilon_{22}=\varepsilon_{33}
=\varepsilon,
\]
\begin{equation}
\mu_{11}=\frac{1}{\varepsilon^3\left(1+\beta^2B_0^4/4\right)},~~~~\mu_{22}=\mu_{33}=\frac{1}{\varepsilon}.
\label{26}
\end{equation}
When the polarization of the electromagnetic wave is parallel to external magnetic field, $\textbf{e}=e_0(1,0,0)$,  we find from Maxwell's equations that $\mu_{22}\varepsilon_{11}\omega^2=k^2$. Then the index of refraction is
\begin{equation}
n_\|=\sqrt{\mu_{22}\varepsilon_{11}}=\sqrt{1+\gamma B_0^2}.
\label{27}
\end{equation}
If the polarization of the electromagnetic wave is perpendicular to external induction magnetic field, $\textbf{e}=e_0(0,1,0)$, we have $\mu_{11}\varepsilon_{22}\omega^2=k^2$, and the index of refraction is given by
\begin{equation}
n_\perp=\sqrt{\mu_{11}\varepsilon_{22}}=\sqrt{\frac{1-\beta^2B_0^4/4}{1+\beta^2B_0^4/4}}.
\label{28}
\end{equation}
The phase velocity depends on the polarization of the electromagnetic wave, and the effect of vacuum birefringence holds. The speed of the electromagnetic wave is $v_\|=1/n_\|$ when the polarization of the electromagnetic wave is parallel to external magnetic field, $\textbf{e}_0\parallel \textbf{B}_0$. In the case $\textbf{e}\perp \textbf{B}_0$, the speed of the electromagnetic wave is $v_\perp=1/n_\perp$.

According to the Cotton-Mouton (CM) effect \cite{Battesti} the coefficient $k_{CM}$ is defined as
\begin{equation}
\triangle n_{CM}=n_\|-n_\perp=k_{CM}B_0^2.
\label{29}
\end{equation}
From Eqs. (27)-(29), using the approximation $ \gamma B_0^2\ll 1$, $\beta B_0^2\ll 1$, we obtain
\begin{equation}
\triangle n_{CM}=\sqrt{1+\gamma B_0^2}-\sqrt{\frac{1-\beta^2B_0^4/4}{1+\beta^2B_0^4/4}}
\approx \frac{\gamma B_0^2}{2},~~~~k_{CM}\approx \frac{1}{2}\gamma.
\label{30}
\end{equation}
Thus, we neglected the terms containing $\beta^2B_0^4$.
The values $k_{CM}$, found in the BMV \cite{Rizzo}, and PVLAS \cite{Valle} experiments are
given by
\[
k_{CM}=(5.1\pm 6.2)\times 10^{-21} \mbox {T}^{-2}~~~~~~~~~~(\mbox {BMV}),
\]
\begin{equation}
k_{CM}=(4\pm 20)\times 10^{-23} \mbox {T}^{-2}~~~~~~~~~~~~~~(\mbox {PVLAS}).
\label{32}
\end{equation}
From Eqs. (30), (31), we evaluate the bounds on the parameter $\gamma$ of our model
\[
\gamma \leq 10^{-20} \mbox {T}^{-2}~~~~~~~~~~(\mbox {BMV}),
\]
\begin{equation}
\gamma \leq 8\times 10^{-23} \mbox {T}^{-2}~~~~~~~~~~(\mbox {PVLAS}).
\label{32}
\end{equation}
It should be noted that the value of $k_{CM}$ calculated from QED is much smaller than the experimental values (31) \cite{Rizzo}, $k_{CM}^{QED}\approx 4.0\times 10^{-24} \mbox {T}^{-2}$. The vacuum birefringence effect
can also be explained by other approaches \cite{Kruglov}, \cite{Kruglov3}, \cite{Kruglov2}, \cite{Kruglov1}.
The improvement in experimental measurements of the vacuum birefringence effect could verify the QED prediction and
possibly to select the electrodynamics model.
Therefore, the model of nonlinear electrodynamics under consideration with two free parameters is of interest.

\section{The energy-momentum tensor and dilatation current}

The symmetrical Belinfante tensor, obtained from Eq. (1) with the help of the method of \cite{Coleman}, is
\begin{equation}
T_{\mu\nu}^{B}=-\frac{1}{\Pi}F_{\nu\alpha}\left(F_{\mu\alpha}-\gamma{\cal G}\tilde{F}_{\mu\alpha}\right)-\delta_{\mu\nu}{\cal L},
\label{33}
\end{equation}
where $\Pi$ is given by Eq. (3). From Eq. (33) one finds the energy density
\begin{equation}
T_{44}^{B}=\frac{1}{\Pi}\left(\textbf{E}^2-\gamma{\cal G}^2\right)+\frac{1}{\beta}\arcsin\left(\beta{\cal F}-\frac{\beta\gamma}{2}{\cal G}^2\right).
\label{34}
\end{equation}
We obtain the trace of the energy-momentum tensor (33):
\begin{equation}
T_{\mu\mu}^{B}=-\frac{4}{\Pi}\left({\cal F}-\gamma{\cal G}^2\right)+\frac{4}{\beta}\arcsin\left(\beta{\cal F}-\frac{\beta\gamma}{2}{\cal G}^2\right).
\label{35}
\end{equation}
The trace of the energy-momentum tensor is not zero \cite{Coleman}, and one finds the dilatation current
\begin{equation}
D_{\mu}^{B}=x_\alpha T_{\mu\alpha}^{B}.
\label{36}
\end{equation}
Then the divergence of dilatation current is given by
\begin{equation}
\partial_\mu D_{\mu}^{B}=T_{\mu\mu}^B.
\label{37}
\end{equation}
So, the scale (dilatation) symmetry is broken as we have introduced the dimensional parameters $\beta$, $\gamma$. The dilatation symmetry is also broken in BI electrodynamics \cite{Kruglov4} but in classical electrodynamics the dilatation symmetry holds.

\subsection{Energy of the point-like charged particle}

Now we calculate the total electric energy of charged point-like particle.
In the case of electrostatics ($\textbf{B}=0$) the electric energy density (34) becomes
\begin{equation}
\rho_E=T^B_{44}=\frac{E^2}{\sqrt{1-\beta^2 E_0^4/4}}-\frac{1}{\beta}\arcsin\left(\frac{\beta E^2}{2}\right).
\label{38}
\end{equation}
Defining the total energy ${\cal E}=\int \rho_E dV$, and using (16), (17), the value $\beta^{1/4}{\cal E}$ is given by
\[
\beta^{1/4}{\cal E}=\frac{e^{3/2}}{4\sqrt{\pi}}\int_0^\infty \left[\frac{\sqrt{2(\sqrt{x^4+1}-x^2)}}{\sqrt{x}}
-\sqrt{x}\arcsin\left(\sqrt{x^4+1}-x^2\right)\right]dx
\]
\begin{equation}
\approx 0.071.
\label{39}
\end{equation}
Implying that the electron mass equals the electromagnetic energy of the point-like charged particle, ${\cal E}=0.51$~MeV, we obtain the parameter $l=\beta^{1/4}=27.6$ fm.
Thus, the old idea of the Abraham and Lorentz \cite{Born1}, \cite{Rohrlich}, \cite{Spohn}, about the electromagnetic
nature of the electron is realized here for the model proposed. At the same time the idea that the electron may be considered classically as a charged object was proposed by Dirac \cite{Dirac1}. We explore this point of view to our model of electrodynamics.

\section{Dirac's quantization}

From Eq. (1), in accordance with the Dirac procedure \cite{Dirac} of gauge
covariant quantization, one obtains the momenta
\begin{equation}
 \pi_i=\frac{\partial {\cal L}}{\partial(\partial_0 A_i)}=-\frac{1}{\Pi}\left(E_i+\gamma{\cal G}B_i\right)=-D_i,~~~
 \pi_0=\frac{\partial {\cal L}}{\partial (\partial_0 A_0)}=0.
\label{40}
\end{equation}
From Eq. (40) we obtain the primary constraint
\begin{equation}
 \varphi_1 (x)\equiv\pi_0 ,\hspace{0.3in}\varphi_1 (x)\approx 0 .
\label{41}
\end{equation}
We explore the Dirac symbol $\approx$ for equations that hold
only weakly. This means that the variable $ \varphi_1 (x)$ may have nonzero Poisson brackets with some quantity. Taking into account the Poisson brackets $\{.,.\}$ between ``coordinates" $A_i(x)$ and momentum $\pi_i(y)$, and the equation $\pi_i=-D_i$, we obtain
\begin{equation}
 \{A_i (\textbf{x},t),D_j(\textbf{y},t)\}=-\delta_{ij}
 \delta(\textbf{x}-\textbf{y}) .
\label{42}
\end{equation}
Applying the operator $\epsilon_{mki}\partial/\partial x_k$ to Eq. (42), one finds the Poisson brackets between $B_m$ and $D_j$
\begin{equation}
 \{B_m (\textbf{x},t),D_j(\textbf{y},t)\}=\epsilon_{mjk}\frac{\partial}{\partial x_k }
 \delta(\textbf{x}-\textbf{y}) .
\label{43}
\end{equation}
In the quantum theory, instead of the Poisson brackets we have to use the quantum commutator. Thus, one should make the replacement $\{B,D\}\rightarrow -i\left[B,D\right]$, where
$\left[B,D\right]=BD-DB$. By virtue of the Lagrangian density (1) and Eq. (40),
and the relation ${\cal H}=\pi_\mu\partial_0 A_\mu-{\cal L}$, we
obtain the Hamiltonian density:
\begin{equation}
 {\cal H}={\bf D}\cdot{\bf E}+\frac{1}{\beta}\arcsin\left(\beta{\cal F}-\frac{\beta\gamma}{2}{\cal G}^2\right)-\pi_m \partial_m A_0 .
\label{44}
\end{equation}
The primary constraint (41) have to be a constant of motion, and we
find the relation
\begin{equation}
 \partial_0 \pi_0 =\{\pi_0,H\}=-\partial_m \pi_m= 0 ,
\label{45}
\end{equation}
where $H=\int d^3x {\cal H}$ is the Hamiltonian. From Eq. (45) we arrive at
the secondary constraint:
\begin{equation}
 \varphi_2 (x)\equiv\partial_m \pi_m ,\hspace{0.3in}\varphi_2 (x)\approx 0.
\label{46}
\end{equation}
Equation (46) represents the Gauss law (see Eq. (9)). One finds the time evolution of the second constraint
\begin{equation}
\partial_0 \varphi_2 =\{\varphi_2,H\}\equiv 0,
\label{47}
\end{equation}
i.e. there are no additional constraints. The second class constraints are absent because
$\{\varphi_1,\varphi_2\}=0$. The similar situation occurs in classical
electrodynamics \cite{Dirac} and in Maxwell's theory on non-commutative spaces \cite{Kruglov6}. In accordance with the Dirac approach \cite{Dirac}, we add to the density of the Hamiltonian the Lagrange multiplier terms $v(x)\pi_0$, $u(x)\partial_m \pi_m$,
\begin{equation}
 {\cal H}_T={\bf D}\cdot{\bf E}+\frac{1}{\beta}\arcsin\left(\beta{\cal F}-\frac{\beta\gamma}{2}{\cal G}^2\right)-\pi_m \partial_m A_0+v(x)\pi_0+u(x)\partial_m \pi_m .
\label{48}
\end{equation}
The functions $v(x)$, $u(x)$ represent auxiliary variables which are connected with gauge degrees of freedom and they do not have physical meaning. Thus, the first class constraints generate the gauge transformations and Eq. (48) is the set of Hamiltonians.
The density energy is given by Eq. (34).
It is convenient to represent the total density of Hamiltonian (48) in terms of fields $A_\mu$ and momenta $\pi_\mu$ to obtain equations of motion. For this purpose we find from Eqs. (3),(7) the electric field $E_i=(\varepsilon^{-1})_{ij}D_j$,
\begin{equation}
\textbf{E}=\Pi\left(\textbf{D}-\frac{\gamma \textbf{B}\cdot \textbf{D}}{1+\gamma \textbf{B}^2}\textbf{B}\right).
\label{49}
\end{equation}
Taking into account Eq. (49), $D_i=-\pi_i$, and $\textbf{B}=\nabla\times\textbf{A}$, one can represent the total Hamiltonian density (48) in terms of fields $A_\mu$ and momenta $\pi_\mu$.
Then we find the Hamiltonian equations
\begin{equation}
\partial_0 A_i=\{A_i,H\}=\frac{\delta H}{\delta \pi_i}= -E_i-\partial_i A_0-\partial_i u(x) ,
\label{50}
\end{equation}
\begin{equation}
\partial_0 \pi_i=\{\pi_i, H\}=-\frac{\delta H}{\delta A_i}=-\nabla\times\textbf{H},
\label{51}
\end{equation}
\begin{equation}
\partial_0 A_0=\{A_0,H\}=\frac{\delta H}{\delta \pi_0}=v(x),\hspace{0.1in}
\partial_0 \pi_0=\{\pi_0,H\}=-\frac{\delta H}{\delta
A_0}=-\partial_m \pi_m . \label{52}
\end{equation}
Eq. (50) is the gauge covariant form of equation for the electric field. Equation (51) is equivalent to the second equation in (9), and the first equation in (9), the Gauss law, is nothing but the second constraint in this Hamiltonian formalism. The function $u(x)$ is arbitrary and one can introduce new function $u'(x)=u(x)+A_0$ so that the $A_0$ is absorbed. After this, the Hamiltonian does not contain the component $A_0$. Therefore, the $A_0$ is not the physical degree of freedom. If we take $v(x)=\partial_0u'(x)$, we obtain from Eqs. (50), (52) the relativistic form of gauge transformations $A'_\mu(x)=A_\mu(x) +\partial_\mu\Lambda(x)$, where $\Lambda(x)=\int dtu'(x)$. In general, there are two arbitrary functions $u'(x)$, $v(x)$. The Hamiltonian equations (50), (51) represent the time evolution of physical fields that are equivalent to the Euler-Lagrange equations.
The first class constraints generate the gauge transformations. Equation (52) represents the time evolution of
non-physical fields because the $A_0$ is arbitrary function connected with the gauge degree of freedom. The variables $\pi_0$, $\partial_m \pi_m$ equal zero as constraints.

In quantum theory, the dynamical variables are $\hat{A}_i$ and $\hat{\pi}_i=-\hat{D}_i$, and they have the commutator \begin{equation}
\left[\hat{A}_i(\textbf{x},t),\hat{D}_j(\textbf{y},t)\right]=-i\delta_{ij}\delta(\textbf{x}-\textbf{y}).
\label{53}
\end{equation}
The wave function $|\Psi\rangle$ obeys the Schr\"{o}dinger equation
\begin{equation}
i\frac{d|\Psi\rangle}{dt}=H|\Psi\rangle,
\label{54}
\end{equation}
where the Hamiltonian $H=\int d^3x {\cal H}$ with the Hamiltonian density (44). According to \cite{Dirac}, the wave function $|\Psi\rangle$ should obey equations as follows
\begin{equation}
\hat{D}_0|\Psi\rangle=0,~~~~\partial_m\hat{D}_m|\Psi\rangle=0.
\label{55}
\end{equation}
Thus, the physical state is invariant under the gauge transformations. In the coordinate representation the operators $\hat{A}_i$, $\hat{\pi}_\mu=-\hat{D}_\mu$ obey the equations
\begin{equation}
\hat{A}_i(x)\Psi[A]=A_i(x)\Psi[A],~~~~\hat{D}_\mu(x)\Psi[A]=i\frac{\delta\Psi[A]}{\delta A_\mu(x)},
\label{56}
\end{equation}
where $\Psi[A]$ is the vector of the state. In this coordinate representation Eqs. (55) become \cite{Matschull}
\begin{equation}
\frac{\delta \Psi[A]}{\delta A_0(x)}=0,~~~~\partial_i\frac{\delta \Psi[A]}{\delta A_i(x)}=0.
\label{57}
\end{equation}
The constraints (55) are restrictions on the state, are the generators of the gauge symmetry, and do not change the physical state $|\Psi\rangle$ which is gauge invariant. The fields $\textbf{E}$, $\textbf{B}$, $\textbf{D}$, $\textbf{H}$ are invariants of the gauge transformations, are measurable quantities (observables), and are represented by the Hermitian operators and do not depend on $A_0$.

\subsection{ The Coulomb Gauge}

Now we consider the gauge fixing method which is beyond the Dirac's approach. With the help of the gauge freedom, described by the functions $v(x)$, $u(x)$, one can impose new constraints
\begin{equation}
\varphi_3 (x)\equiv A_0\approx 0 ,\hspace{0.3in}\varphi_4 (x)\equiv\partial_m
A_m\approx 0 .
\label{58}
\end{equation}
As a result, equations occur \cite{Hanson}
\begin{equation}
\{\varphi_1 (\textbf{x},t),\varphi_3
(\textbf{y},t)\}=-\delta(\textbf{x}-\textbf{y}) ,\hspace{0.1in}
\{\varphi_2 (\textbf{x},t),\varphi_4
(\textbf{y},t)\}=\Delta_x\delta(\textbf{x}-\textbf{y}) ,
\label{59}
\end{equation}
where $\Delta_x\equiv\partial^2/(\partial x_m)^2$. We may define the
``coordinates" $Q_i$ and the conjugated momenta $P_i$ ($i=1,2$),
\begin{equation}
Q_i=(A_0,\partial_m
A_m),\hspace{0.3in}P_i=(\pi_0,-\Delta^{-1}_x\partial_m \pi_m) ,
\label{60}
\end{equation}
so that
\begin{equation}
\{Q_i(\textbf{x},t),P_j(\textbf{y},t)\}=\delta_{ij}
\delta(\textbf{x}-\textbf{y}) ,~~
\Delta^{-1}_x=-\frac{1}{4\pi|\textbf{x}|} ,~~\Delta_x
\frac{1}{4\pi|\textbf{x}|}=-\delta(\textbf{x}).
\label{61}
\end{equation}
The canonical variables $Q_i,P_i$ in Eqs. (60) are not the physical
degrees of freedom and must be eliminated. By virtue of the
Poisson brackets matrix \cite{Hanson} $C_{ij}=\{\varphi_i
(\textbf{x},t),\varphi_j (\textbf{y},t)\}$, and the definition of
the Dirac brackets \cite{Hanson}, \cite{Henneaux}, one finds
\begin{equation}
\{\pi_0(\textbf{x},t),A_0(\textbf{y},t)\}^*=
\{\pi_0(\textbf{x},t),A_i(\textbf{y},t)\}^*=
\{\pi_i(\textbf{x},t),A_0(\textbf{y},t)\}^*=0 ,
\label{62}
\end{equation}
\begin{equation}
\{\pi_i(\textbf{x},t),A_j(\textbf{y},t)\}^*=
-\delta_{ij}\delta(\textbf{x}-\textbf{y})+\frac{\partial^2}{\partial
x_i \partial y_j} \frac{1}{4\pi|\textbf{x}-\textbf{y}|}
\hspace{0.2in}(i,j=1,2,3) ,
\label{63}
\end{equation}
\begin{equation}
\{\pi_\mu(\textbf{x},t),\pi_\nu(\textbf{y},t)\}^*=
\{A_\mu(\textbf{x},t),A_\nu(\textbf{y},t)\}^*=0
\hspace{0.3in}(\mu, \nu=1,2,3,4) .
\label{62}
\end{equation}
Making the Fourier transformation, Eq. (63) becomes
\begin{equation}
\{\pi_i(\textbf{k}),A_j(\textbf{q})\}^*=-(2\pi)^3
\delta(\textbf{k+q})\left(\delta_{ij}-\frac{k_i
k_j}{\textbf{k}^2}\right) .
\label{65}
\end{equation}
In the right side of Eq. (65) the projection operator $\Pi_{ij}=\delta_{ij}-k_i
k_j/\textbf{k}^2$ extracts the physical transverse components of vectors.

One may put all second class constraints to be zero, and only transverse
components of the vector potential $A_\mu$ and the momentum $\pi_\mu$
are physical independent variables. The operators (60) will be
absent in the reduced physical phase space and the physical
Hamiltonian is $H^{ph}=\int d^3 x T^B_{44}$.
Replacing the Poisson brackets by the Dirac
brackets we obtain from this Hamiltonian equations of
motion whch coincide with Eqs. (50)-(52) at $u(x)=v(x)=0$. We have to substitute Dirac's
brackets, in the quantum theory, by the quantum commutators,
$\{.,.\}^*\rightarrow-i[.,.]$. In this approach, only physical degrees of freedom remain in the theory.

To quantize the theory, one can use the functional integral approach. In this case, to take into consideration the gauge degrees of freedom, we should insert a gauge condition in the functional integral. Then one needs to introduce
ghosts. Another way is to explore the Fock basis without introducing the wave functionals. Then one can construct the vacuum state and the scalar product of wave functions on the physical state space. In this procedure, we should introduce negative norm states that have not to appear in the physical spectrum.

\section{Conclusion}

We have proposed a new model of nonlinear electrodynamics possessing two dimensional parameters $\beta$, $\gamma$
and obtained the corrections to Coulomb's law. There is the effect of vacuum birefringence if the external constant and uniform induction magnetic field is present. If $\gamma=0$ the phenomenon of vacuum birefringence vanishes. The bounds on the value $\gamma$ were estimated from the data of BMV and PVLAS experiments, $\gamma \leq 10^{-20}$ T$^{-2}$ (BMV), $\gamma \leq 8\times 10^{-23}$ T$^{-2}$ (PVLAS). The effect of vacuum birefringence predicted by QED was not conformed by the experiment, and therefore, models
of electrodynamics which may not be consistent with QED in this case are of interest. 

We have calculated the indices of refraction for two polarizations of electromagnetic waves, parallel and perpendicular to the magnetic field so that phase velocities of electromagnetic waves depend on polarizations. The symmetrical Belinfante energy-momentum tensors and dilatation current were found and we show that the dilatation symmetry is violated. The scale symmetry is broken as the dimensional parameters $\beta$, $\gamma$ are introduced. The electric field of a point-like charge is finite at the origin in the model considered. We have calculated the finite electromagnetic energy of point-like charged particles. For $l\equiv\beta^{1/4}=27.6$ fm the electron mass equals the total electromagnetic energy and we can assume that the mass of the electron has a pure electromagnetic nature. We have considered the gauge covariant quantization of the nonlinear electrodynamic fields which is similar to the quantization of Maxwell's fields. The gauge fixing approach based on Dirac's brackets has also studied. It is interesting to investigate the interactions between electromagnetic field described
by arcsin-electrodynamics and fermion fields. We leave such study for further consideration.

\end{document}